\documentclass[aps,prc,twocolumn,groupedaddress,showpacs,eqsecnum]{revtex4}
\usepackage{graphicx}
\usepackage{dcolumn}
\usepackage{bm}

\newcommand{\raa}{($\alpha$,$\alpha$)}

\newcommand{\rag}{($\alpha$,$\gamma$)}

\newcommand{\heaa}{$^{3}$He\raa $^{3}$He}
\newcommand{\heag}{$^{3}$He\rag $^{7}$Be}

\newcommand{\sfac}{S-factor}

\newcommand{\al}{$\alpha$}

\begin{document}

\title{
Low-energy $^3$He($\alpha$,$\alpha$)$^3$He elastic scattering and the
$^3$He($\alpha$,$\gamma$)$^7$Be reaction
}

\author{Peter Mohr  
}
\email[E-mail: ]{WidmaierMohr@t-online.de}
\affiliation{
Diakonie-Klinikum Schw\"abisch Hall, D-74523 Schw\"abisch Hall,
Germany
}

\date{\today}

\begin{abstract}
The cross sections of the $^3$He($\alpha$,$\alpha$)$^3$He and
$^3$He($\alpha$,$\gamma$)$^7$Be reactions are studied at low energies using a
simple two-body model in combination with a double-folding potential. At very
low energies the capture cross section is dominated by direct $s$-wave
capture. However, at energies of several MeV the $d$-wave contribution
increases, and the theoretical capture cross section depends sensitively on
the strength of the $L=2$ potential. Whereas the description of the $L=2$
elastic phase shift requires a relatively weak potential strength, recently
measured capture data can only be described with a significantly enhanced
$L=2$ potential. A simultaneous description of the new experimental capture
data and the elastic phase shifts is not possible within this model. Because
of the dominating extranuclear capture, this conclusion holds in general for
most theoretical models.
\end{abstract}


\maketitle

\section{Introduction}
\label{sec:intro}
The \heag\ capture reaction is a key reaction in nuclear astrophysics. The
flux of solar neutrinos at higher energies depends on the branching between
the $^3$He($^3$He,2p)$\alpha$ and \heag\ reactions, and in big-bang
nucleosynthesis $^7$Li can be produced via the \heag\ capture reaction and
subsequent electron capture of $^7$Be \cite{Cyb04}. While the energy range
of big-bang nucleosynthesis is covered by experimental data, the experiments
are approaching the Gamow window around $E_0 = 22$\,keV of the \heag\ reaction
in the sun, but still have not reached $E_0$.

Very recently, it has been attempted successfully to extend the measured
energy range of the \heag\ reaction to higher energies up to about 3.2\,MeV
\cite{Lev09}. From these new experimental data the energy dependence of the
\heag\ capture reaction can be extracted and compared to theoretical
predictions. A better theoretical understanding of the energy dependence will
help to reduce the uncertainties of the extrapolation of the cross section
down to the Gamow window at the temperature of the interior of our
sun. Several further experimental data sets exist at lower energies
\cite{Hol59,Par63,Nag69,Kra82,Rob83,Volk83,Alex84,Osb84,Hil88,Nar04,Bemm06,Gyu07,Bro07,Con07,Cos08}
that are summarized in two recent compilations \cite{SolFus,NACRE}.

Besides theoretical calculations shown in the experimental papers
\cite{Hol59,Par63,Nag69,Kra82,Rob83,Volk83,Alex84,Osb84,Hil88,Nar04,Bemm06,Gyu07,Bro07,Con07,Cos08},
various theoretical studies have been devoted to the analysis of the
\heag\ capture cross section
\cite{Chr61,Tom61,Tom63,Kim81,Liu81,Wal83,Kaj84,Wal84,Mer86,Kaj87,Buck88,Mohr93,Dub95,Bay00,Cso00,Nol01,Des04,Can08,Mas09,Nav09}. Additionally,
a review on the status of the \heag\ reaction is given in \cite{Cyb08}.

Theoretical models can be significantly constrained by the request that
elastic scattering data have to be described simultaneously with the capture
data. Phase shifts have been derived from elastic scattering angular
distributions in \cite{Spi67,Boy72,Har72}, and angular distributions or
excitation functions at low energies are reported in
\cite{Bar64,Iva68,Chu71,Mohr93}. 

It will be shown that the capture cross section of the \heag\ reaction at low
energies is dominated by E1 transitions from incoming $s$- and $d$-waves to
bound $p$-states in $^7$Be. Consequently, a precise description of the
$s$-wave and $d$-wave phase shifts is a prerequisite for the calculation of
the \heag\ capture cross section.

This article is organized as follows. In Sec.~\ref{sec:DC} a brief
description of the direct capture model is provided. General remarks on the
applicability and reliability of this model and other theoretical calculations
are given in Sec.~\ref{sec:gen}. Results for the \heaa\ elastic scattering
and \heag\ capture cross sections are presented in Sec.~\ref{sec:res}, and
theoretical uncertainties are carefully analyzed. Conclusions are drawn in
Sec.~\ref{sec:conc}, and finally a brief summary is given in
Sec.~\ref{sec:summ}. 

All energies $E$ are given in the center-of-mass system throughout this paper
except explicitly noted. Excitation energies $E^\ast$ in $^7$Be are related to
$E$ by $E^\ast = E + Q$ with the $Q$-value of the \heag\ reaction $Q =
1586.6$\,keV \cite{Til02}.

\section{The direct capture model}
\label{sec:DC}
The direct capture (DC) model is a simple but powerful model to calculate
capture cross sections between light nuclei. The full formalism is given
explicitly in \cite{Kim87}, and its application together with systematic
folding potentials is described in detail in our previous work on the
\heaa\ and \heag\ reactions \cite{Mohr93}. Here I only repeat some essential
features of the DC model that are important for the following discussion.

\begin{figure}[htbp] 
  \centering
  \includegraphics[bbllx=210,bblly=85,bburx=330,bbury=315,width=4cm]{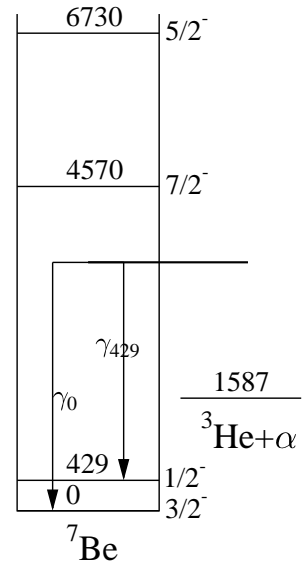}
  \caption{
   Simplified level scheme of $^7$Be with the $L=1$ and $L=3$ cluster
   states. Excitation energies $E^\ast$ are given in keV. All data are taken
   from \cite{Til02}.
}
  \label{fig:a7level}
\end{figure}

The theoretical capture cross section $\sigma_{\rm{th}}$ is given by the
product of the DC cross section $\sigma_{\rm{DC}}$ and the spectroscopic
factor $C^2 S$ of the final state:
\begin{equation}
\sigma_{\rm{th}} = C^2 S \, \times \, \sigma_{\rm{DC}}
\label{eq:sig_th}
\end{equation}
The DC cross section depends on the square of the overlap between the
scattering wave function $\chi_{L_i,J_i}(r)$, the bound state wave function
$u_{N,L_f,J_f}(r)$, and the electromagnetic operator ${\cal{O}}^{E1,E2,M1}$ of
$E1$, $E2$, and $M1$ transitions:
\begin{equation}
\sigma_{\rm{DC}} \sim 
\Bigl| \int \chi_{L_i,J_i}(r) \, {\cal{O}}^{E1,E2,M1} \, u_{N,L_f,J_f}(r) \,
dr \Bigr|^2
\label{eq:DC}
\end{equation}
where $L_i,J_i$ and $L_f,J_f$ are the angular momenta in the initial
scattering wave function $\chi(r)$ and the final bound state wave function
$u(r)$. $N$ is the number of nodes in the bound state wave function that
takes into account the Wildermuth condition. The total number of oscillator
quanta $Q = 2N + L_f = 3$ for three nucleons in the $1p$ shell leads to two
cluster states in $^7$Be. There is a first state with one node ($N=1$) and
angular momentum $L=1$, and a second state without node ($N=0$) and
$L=3$. Both states are split into dubletts because of the spin $S = 1/2$ of
the $^3$He nucleus. Both $L=1$ states are located below the $^3$He-$\alpha$
threshold in $^7$Be, whereas the $L=3$ states are located above the threshold
and may appear as resonances in the \heag\ capture reaction. Properties of the
cluster states are given in Table \ref{tab:level}, and a simplified level
scheme of $^7$Be is shown in Fig.~\ref{fig:a7level}.
\begin{table}[htbp]
\caption{
Level scheme of $^7$Be. Experimental data are taken from \cite{Til02}. The
potential strength parameters are adjusted to reproduce the energies of the
bound and quasi-bound states (see text).
\label{tab:level}
}
\begin{center}
\begin{tabular}{ccc@{$\pm$}crc@{$\pm$}cc}
$L$	& $J^{\pi}$ 
	& \multicolumn{2}{c} {$E^\ast_{\rm exp}$} 
	& $E_{\rm{exp}} = E_{\rm calc}$
	& \multicolumn{2}{c} {$\Gamma_{\rm exp}$} 
	& $\Gamma_{\rm calc}$ \\
	& 
	& \multicolumn{2}{c} {$({\rm keV})$} 
	& \multicolumn{1}{c} {$({\rm keV})$} 
	& \multicolumn{2}{c} {$({\rm keV})$} 
	& $({\rm keV})$ \\
\hline
1 & $3/2^-$ & \multicolumn{2}{c} {0} & $-1586.6$\footnote{potential adjusted}
	& \multicolumn{2}{c} {($T_{1/2} = 53.22~\rm{d}$)}
		& -- \\
1 & $1/2^-$ & 429.1 & 0.1 & $-1157.5$\footnotemark[1]
& \multicolumn{2}{c} {($\tau = 192 \pm 25~\rm{fs}$)}
		& -- \\
3 & $7/2^-$ & 4570 & 50	& 2983.4\footnotemark[1]
	& \multicolumn{2}{c} {$175 \pm 7$}
		& 145 \\
3 & $5/2^-$ & 6730 & 100	& 5143.4\footnotemark[1]
	& \multicolumn{2}{c} {$\approx 1200$}
		& $\approx 1100$ \\
\hline
\end{tabular}
\end{center}
\end{table}

\begin{figure}[thbp] 
  \centering
  \includegraphics[bbllx=55,bblly=95,bburx=320,bbury=360,width=8.5cm]{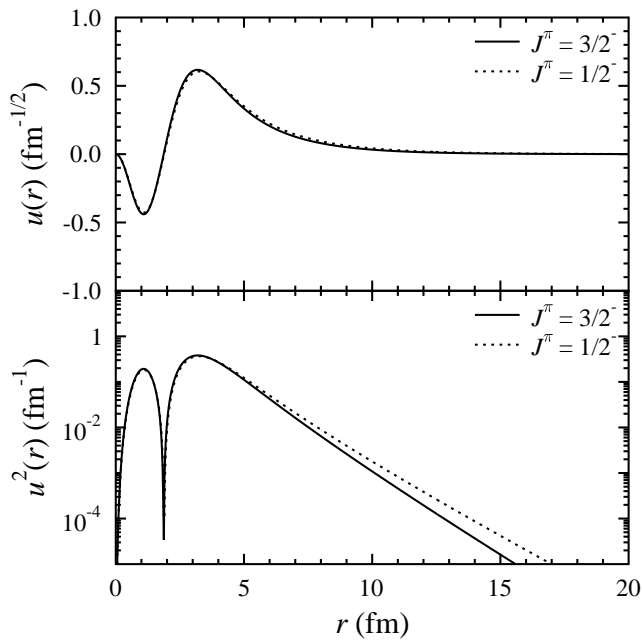}
  \caption{
   Bound state wave functions $u(r)$ for the $L = 1$, $J^\pi = 3/2^-$ ground
   state and the $L = 1$, $J^\pi = 1/2^-$ first excited state in $^7$Be in
   linear scale (upper) and $u^2(r)$ in logarithmic scale (lower). The wave
   functions are very similar in the nuclear interior. The slope in the
   exterior reflects the different binding energies ($E_B = -1587$\,keV for
   the $3/2^-$ ground state and $E_B = -1158$\,keV for the $1/2^-$ first
   excited state).
}
  \label{fig:bound}
\end{figure}

\begin{figure}[thbp] 
  \centering
  \includegraphics[bbllx=55,bblly=95,bburx=320,bbury=475,width=8.5cm]{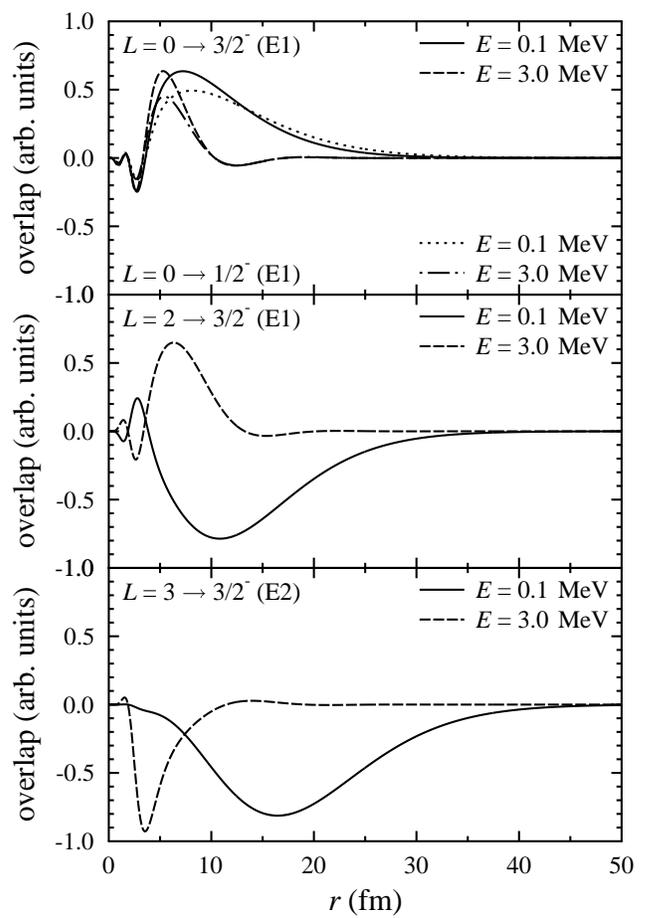}
  \caption{
   Overlap of the scattering wave function $\chi(r)$, electromagnetic
   transition operator $\cal{O}$, and bound state wave function $u(r)$ for
   different transitions of the \heag\ capture reaction. Upper: Non-resonant
   $E1$ $s$-wave capture to the $3/2^-$ ground state and $1/2^-$ first excited
   state at energies $E = 0.1$\,MeV and 3.0\,MeV. Middle: Non-resonant $E1$
   $d$-wave capture to the $3/2^-$ ground state at $E = 0.1$\,MeV and
   3.0\,MeV. Lower: Non-resonant and resonant $E2$ $f$-wave capture to the
   $3/2^-$ ground state at $E = 0.1$\,MeV and 3.0\,MeV. Further discussion see
   text.
}
  \label{fig:overlap}
\end{figure}

The bound state wave functions $u(r)$ are shown for the $L = 1$, $J^\pi =
3/2^-$ ground state and the $L = 1$, $J^\pi = 1/2^-$ first excited state in
$^7$Be in Fig.~\ref{fig:bound}. Fig.~\ref{fig:overlap} shows the integrand of
Eq.~(\ref{eq:DC}), i.e.\ the overlap of the scattering wave function
$\chi(r)$, the electromagnetic operator $\cal{O}$, and the bound state wave
function $u(r)$ for different transitions in the \heag\ capture reaction at a
very low (100\,keV) and at a higher energy (3\,MeV). A detailed discussion
will be given in the following Sec.~\ref{sec:gen}.

The basic ingredient for the calculation of the DC integral in
Eq.~(\ref{eq:DC}) are the potentials for the entrance channel (elastic
scattering) and the exit channel (bound state wave function). As soon as the
potentials are fixed, the DC integral is calculated without further adjustment
of parameters to experimental capture data. The spectroscopic factor $C^2 S$
of the bound state is taken -- exactly as in \cite{Mohr93} -- from theory in
the present study: $C^2 S(3/2^-) = 1.174$ and $C^2 S(1/2^-) = 1.175$
\cite{Kur75}. The spectroscopic factor is considered as an absolute
normalization for the calculated capture cross section, similar to the
procedure in \cite{Ili08}. It is beyond the scope of the present paper to
discuss the relation between the asymptotic normalization coefficient and the
spectroscopic factor as e.g.\ in \cite{Muk08} because the main conclusions of
this work are not affected.

The potential for the $^3$He $-$ \al\ system is the sum of the central nuclear
potential, the spin-orbit potential, and the Coulomb potential. The real part
of the central nuclear potential is calculated from the double folding
procedure \cite{Sat79,Kob84} that is scaled by a strength parameter $\lambda$
which is about $1.4 - 1.8$ in this study. The underlying nuclear densities are
derived from the measured charge density distributions \cite{Vri87}. The
imaginary part of the nuclear potential can be set to zero because there are
no open channels below 4\,MeV except the relatively weak \heag\ capture. The
spin-orbit potential is taken in the usual Thomas form proportional to $1/r \,
\times \, dV/dr$, again scaled by a spin-orbit strength parameter
$\lambda_{LS}$. Finally, the Coulomb potential $V_C$ is calculated from the
homogeneous charged sphere model with a Coulomb radius $R_C$ identical to the
root-mean-square radius of the folding potential: $R_C = r_{\rm{rms}} =
2.991$\,fm. Further details can be found in \cite{Mohr93}. The total potential
is given by:
\begin{equation}
V(r) = \lambda \, V_F(r) 
+ \lambda_{LS} \, \frac{\rm{fm}^2}{r} \, \frac{dV_F(r)}{dr} \, \vec{L} \vec{S} 
+ V_C(r)
\label{eq:pot}
\end{equation}
with the unscaled ($\lambda = 1$) folding potential $V_F(r)$. The factor
fm$^2$ in the spin-orbit potential is added to obtain a dimensionless strength
parameter $\lambda_{LS}$. $\vec{L}$, $\vec{S}$, and $\vec{J} = \vec{L} +
\vec{S}$ are the orbital, spin, and total angular momenta in units of $\hbar$.

One main advantage of folding potentials is the small number of adjustable
parameters. The shape of the potential is fixed by the folding procedure. Only
the strength parameters $\lambda$ and $\lambda_{LS}$ have to be adjusted to
experimental data. Obviously, the small number of adjustable parameters
improves the predictive power of the calculations.

The shape of the folding potential for $^3$He $-$ \al\ is almost
Gaussian. Consequently, the spin-orbit part has almost the same shape, and the
sum of central and spin-orbit potential again has the same shape. In other
words, the spin-orbit potential slightly increases the potential for $J = L +
1/2$ waves and slightly decreases the potential for the $J = L - 1/2$ waves
but keeps the shape of the folding potential. Alternatively, the same effect
can be obtained if the spin-orbit potential is set to zero and instead the
strength parameter of the central potential becomes $J$-dependent. This
approach has been followed e.g.\ in \cite{Dub95}: there very similar results
to our previous work \cite{Mohr93} were obtained using empirical Gaussian
potentials with a $J$-dependent depth $V_0$.

\section{General remarks}
\label{sec:gen}
The nucleus $^7$Be and the \heaa\ elastic scattering and \heag\ capture
reactions are textbook examples for the successful application of a simple
two-body model because of the strong internal binding energies of the two
constituents $^3$He and $^4$He. Early calculations have been performed for the
\heag\ reaction using hard-sphere phase shifts for the entrance channel, thus
considering external capture only. Already these early calculations have
successfully reproduced the cross section of the \heag\ capture reaction at
low energies \cite{Chr61,Tom61,Tom63}. Nowadays it must be the aim of
theoretical studies to reproduce simultaneously the \heaa\ elastic scattering
cross sections and phase shifts, the \heag\ capture reaction, and
electromagnetic properties of the $^7$Be nucleus. Often also the mirror
nucleus $^7$Li and the mirror reaction $^3$H\rag $^7$Li are considered. The
present study focuses on the new results for the \heag\ reaction \cite{Lev09};
the $^7$Li mirror system was already studied in our earlier work
\cite{Mohr93}.

Only very few radiation widths $\Gamma_\gamma$ or lifetimes $\tau$ have been
measured for the $^7$Be nucleus. The first excited $1/2^-$ state in $^7$Be
decays by a $M1$ transition to the $^7$Be ground state with a lifetime of
$\tau_{\rm{m}} = 192 \pm 25$\,fs \cite{Til02} corresponding to $B(M1,3/2^-
\rightarrow 1/2^-) = 1.87 \pm 0.25\,\mu_N^2$. Following the formalism in
\cite{Mas09,Wal85,Buck85}, the reduced transition strength $B(M1,\,3/2^-
\rightarrow 1/2^-)$ is given by
\begin{equation}
B(M1) = \frac{1}{4\pi} \left[ 2 \mu(^3{\rm{He}}) -
\mu_N G \right]^2 \bigl| < u_{3/2^-} | u_{1/2^-} > \bigr|^2
\label{eq:BM1}
\end{equation}
with the effective cluster-cluster orbital gyromagnetic factor $G = (Z_1 A_2^2
+ Z_2 A_1^2) / [ A_1 A_2 (A_1 + A_2)] = 0.5952$ for $^7$Be = $^3$He $\otimes$
\al\ and $\mu(^3{\rm{He}}) = -2.1276\,\mu_N$ \cite{Kel87}. As already
pointed out in \cite{Buck88}, the overlap of the bound state wave functions
$\bigl| < u_{3/2^-} | u_{1/2^-} > \bigr|^2$ is close to unity because both
wave functions $u(r)$ are very similar. This similarity is also obvious for
the wave functions in this study, see Fig.~\ref{fig:bound}; here the overlap
deviates by less than one per cent from unity. Consequently, the reduced
transition strength $B(M1)$ is practically defined by the factor
$\frac{1}{4\pi} \left[ 2 \mu(^3{\rm{He}}) - \mu_N G \right]^2 =
1.873\,\mu_N^2$ which is in excellent agreement with the experimental result
of $1.87 \pm 0.25\,\mu_N^2$. The excellent reproduction of the $B(M1)$
transition strength between the $3/2^-$ ground state and the $1/2^-$ first
excited state in $^7$Be is thus a more or less trivial result for any two-body
calculation with realistic bound state wave functions.

The $E2$ contribution of this transition
is orders of magnitude smaller than the $M1$ transition because of the
relatively low transition energy of $E_\gamma = 429$\,keV and the strong
$E_\gamma^5$ dependence of $E2$ transitions: one $E2$ Weisskopf unit
corresponds to a radiation width of about $\Gamma_\gamma(E2) \approx 10$\,neV
which has to be compared to the experimental $M1$ width of $\Gamma_\gamma =
3.4$\,meV. 

No further experimental data for radiation widths in $^7$Be exist in
\cite{Til02}. However, the recent experiment \cite{Lev09} has measured the
cross section of the \heag\ reaction in the $7/2^-$ resonance for the first
time, and it was possible to derive the radiation width $\Gamma_{\gamma,0}$ in
\cite{Lev09}. It will be shown in Sec.~\ref{sec:rescapt} that the DC model is
able to reproduce the experimental value within its relatively large
experimental uncertainty. Contrary to the above studied $M1$ transition, this
result is not trivial because it requires the overlap of the $7/2^-$
scattering wave function, the $E2$ operator, and the $3/2^-$ ground state wave
function (see Fig.~\ref{fig:overlap}, lower part).

Often, ``direct capture'' is also named ``extranuclear capture'' or
``non-resonant capture''. The origin of these somewhat misleading names comes
from the early calculations by Christy and Duck \cite{Chr61} and Tombrello and
coworkers \cite{Tom61,Tom63}. A good description of the experimental
\heag\ capture cross section was obtained in \cite{Chr61,Tom61,Tom63} using
hard-sphere phase shifts for the non-resonant $s$- and $d$-waves (that are by
definition non-resonant). As can be seen from Fig.~\ref{fig:overlap} (upper
and middle parts), the main contribution of the $E1$ transitions from incoming
$s$-waves and $d$-waves to the bound $L=1$ states comes indeed from the
nuclear exterior, thus validating the hard-sphere approximation for the
$s$-waves and $d$-waves in
\cite{Chr61,Tom61,Tom63}. However, as can be seen from Fig.~\ref{fig:overlap}
(lower part), the main contribution of the $E2$ transition from the incoming
$f$-wave to the $3/2^-$ ground state comes from the exterior at the low energy
of $E = 100$\,keV, but is shifted to the interior at $E = 3.0$\,MeV which is
in the $7/2^-$ resonance. Thus, the terminus ``direct capture'' is neither
identical to ``extranuclear capture'' nor to ``non-resonant capture''. This
has also been illustrated for the $3^+$ resonance in the $^2$H\rag $^6$Li
reaction in \cite{Mohr94}.

\section{Results}
\label{sec:res}

\subsection{Elastic scattering cross sections and phase shifts}
\label{sec:resscat}
\begin{figure}[thbp] 
  \centering
  \includegraphics[bbllx=30,bblly=95,bburx=290,bbury=735,width=8.0cm]{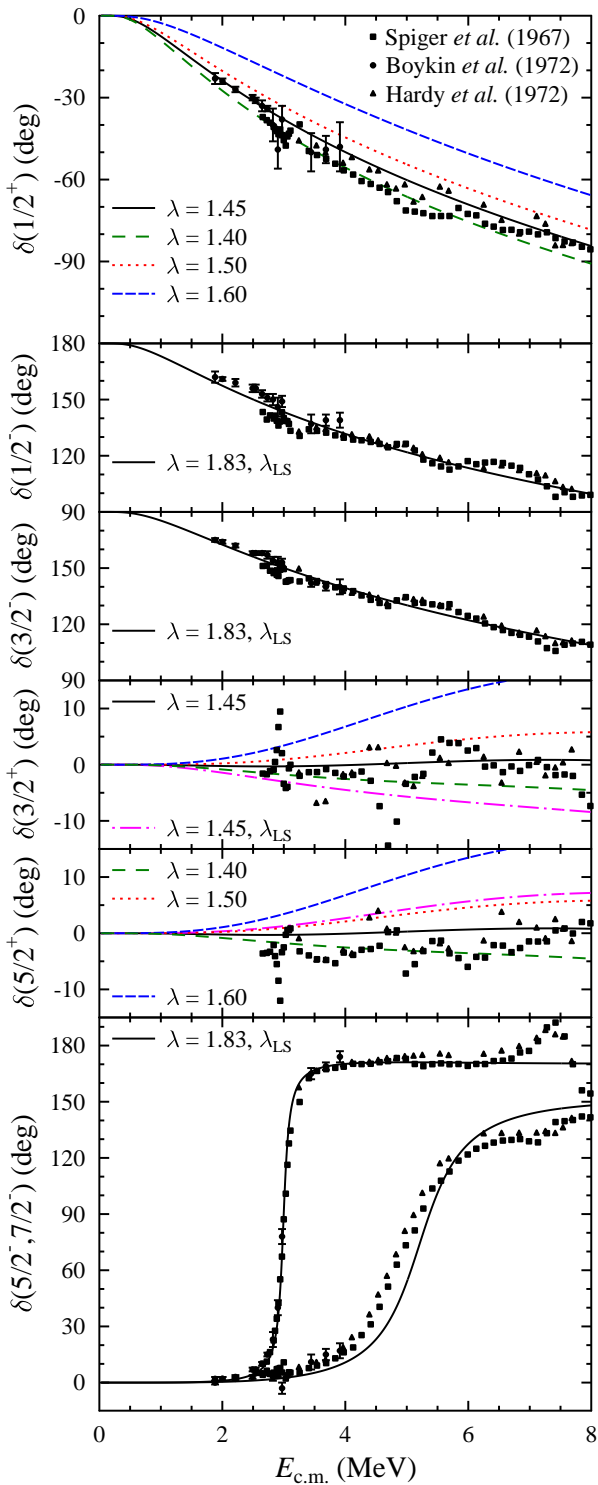}
  \caption{
    (Color online)
    Phase shifts for elastic \heaa\ scattering. Experimental data are taken
    from \cite{Spi67,Boy72,Har72}. Calculations are performed using the
    potential strength parameters $\lambda^{\rm{even}} = 1.452$,
    $\lambda_{LS}^{\rm{even}} = 0$, $\lambda^{\rm{odd}} = 1.830$ and
    $\lambda_{LS}^{\rm{odd}} = -0.173$ (standard case, full black line). The
    parameter $\lambda^{\rm{even}}$ is varied between 1.40 and 1.60
    ($\lambda^{\rm{even}} = 1.40$: green dashed line; $\lambda^{\rm{even}} =
    1.50$: red dotted line; $\lambda^{\rm{even}} = 1.60$: blue narrow-dashed
    line), and the spin-orbit strength of the odd partial waves has also been
    used for the even partial waves ($\lambda^{\rm{even}} = 1.452$,
    $\lambda_{LS}^{\rm{even}} = -0.173$: magenta dash-dotted line). Further
    discussion see text.
}
  \label{fig:a7phase}
\end{figure}

As already pointed out above, the parameters of the potential -- i.e.\ the
potential strength parameters $\lambda$ and $\lambda_{LS}$ -- have to be
adjusted to experimental data. Elastic scattering phase shifts are typically
used for this adjustment. Such phase shifts have been derived from angular
distributions and excitation functions over a broad range of energies
\cite{Spi67,Boy72,Har72}, see Fig.~\ref{fig:a7phase}.

The data of Spiger and Tombrello cover the energy range slightly below the
$7/2^-$ resonance and range from about 2.5\,MeV up to 10\,MeV
\cite{Spi67}. The data of Boykin {\it et al.}\ cover the energy range from
about 1.9\,MeV up to 4\,MeV. However, no numerical results are given for the
important $d$-waves. Instead, it is pointed out that ``the values obtained at
the various energies were scattered with no discernible trend in a band
between $-4^\circ$ and $+4^\circ$'' \cite{Boy72}. The data from Hardy {\it et
  al.}\ start above the $7/2^-$ resonance and range from about 3.3\,MeV up to
7.7\,MeV \cite{Har72}.

The adjustment of the potential strength parameters $\lambda$ and
$\lambda_{LS}$ is done as follows. The procedure is almost identical to the
previous work \cite{Mohr93} with a modification of the spin-orbit
potential. First, the strength parameter $\lambda$ is adjusted to the $s$-wave
phase shift. This can be done unambiguously because the $s$-wave is not
affected by the spin-orbit potential. An excellent agreement with the
experimental phase shifts is obtained with $\lambda(L=0) = 1.452$ (full black
line in Fig.~\ref{fig:a7phase}, upper part). Additionally, the results are
shown for a variation of the strength parameter $\lambda$ between $1.4$ and
$1.6$ (green dashed, red dotted, and blue narrow-dashed lines) that will be
important for the analysis of the \heag\ capture cross section in
Sec.~\ref{sec:rescapt}.

The parameter $\lambda(L=0) = 1.452$ is also used for the $d$-wave phase
shifts. As can be seen from Fig.~\ref{fig:a7phase}, there is excellent
agreement with the experimental $d$-wave phase shifts over a broad energy
range. There is no evidence for a different behavior of the $d_{3/2}$ and
$d_{5/2}$ phase shifts from the data in \cite{Spi67,Boy72,Har72}. Thus, the
spin-orbit potential has to vanish for the $d$-waves, and $\lambda_{LS}(L=2) =
0$. It is interesting to note that any deviation of the potential strength
from $\lambda = 1.452$ leads to deviations of the calculated $d$-wave phase
shifts from the experimentally vanishing data. Combining the results for
the $s$-wave and the $d$-waves, the potential strength parameters for even
partial waves are $\lambda^{\rm{even}} = 1.452$ and $\lambda_{LS}^{\rm{even}}
= 0$. A test with the spin-orbit strength derived from the $f$-waves (see
below) also shows clear disagreement to the experimental $d$-wave phase shifts
(magenta dash-dotted line in Fig.~\ref{fig:a7phase}).

A noticeably higher value for the potential strength $\lambda$ is found for
the odd partial waves. $\lambda^{\rm{odd}}$ and $\lambda_{LS}^{\rm{odd}}$ have
been adjusted to reproduce the energies of the $f$-wave resonances (see
Fig.~\ref{fig:a7phase}). This leads to $\lambda^{\rm{odd}} = 1.830$ and
$\lambda_{LS}^{\rm{odd}} = -0.173$. Using these values, an excellent agreement
for the $f$-wave phase shifts is obtained in both non-resonant and resonant
energy regions. In particular, the resonance widths are properly reproduced
(see also Table \ref{tab:level}). Good agreement with the experimental
$p$-wave phase shifts is also obtained with the above $\lambda^{\rm{odd}}$ and
$\lambda_{LS}^{\rm{odd}}$. The parameters $\lambda^{\rm{odd}}$ and
$\lambda_{LS}^{\rm{odd}}$ were not varied in the following study because any
change of these parameters shifts the $f$-wave resonances dramatically in
energy.

\begin{figure}[thbp] 
  \centering
  \includegraphics[bbllx=95,bblly=130,bburx=345,bbury=605,width=8.5cm]{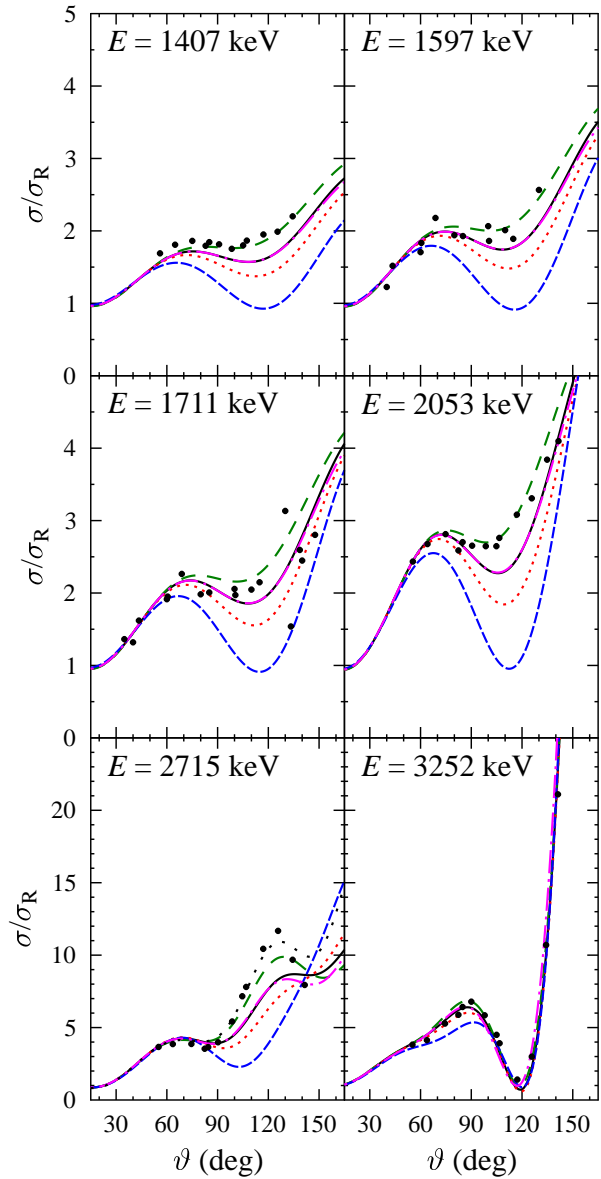}
  \caption{
   (Color online)
   Angular distributions of \heaa\ elastic scattering, normalized to
   Rutherford scattering. Experimental data are
   taken from \cite{Bar64,Mohr93}. The linestyle and color
   coding is identical to Fig.~\ref{fig:a7phase}; the dotted black line at $E
   = 2715$\,keV shows the artificial enhancement of the contribution of the
   $7/2^-$ resonance (see text).
}
  \label{fig:scatt}
\end{figure}

The new experimental \heag\ capture data cover an energy range that is at the
lower end of the experimental phase shift data of
\cite{Spi67,Boy72,Har72}. Fortunately, further experimental angular
distributions of \heaa\ elastic scattering are available in
\cite{Bar64,Chu71,Mohr93} in the relevant energy range of the new
\heag\ experiment \cite{Lev09}.  These experimental angular distributions are
compared to the calculated cross sections using the above determined
parameters $\lambda^{\rm{even}} = 1.452$, $\lambda_{LS}^{\rm{even}} = 0$,
$\lambda^{\rm{odd}} = 1.830$ and $\lambda_{LS}^{\rm{odd}} = -0.173$. The
result is shown in Fig.~\ref{fig:scatt} (full black lines). Additionally,
again the sensitivity on the potential strength parameters
$\lambda^{\rm{even}}$ and $\lambda_{LS}^{\rm{even}}$ is analyzed (colored
dashed and dotted lines in Fig.~\ref{fig:scatt}).

The conclusions from Fig.~\ref{fig:scatt} are clear. As expected, an excellent
agreement between the experimental data and the calculations is only obtained
using the same parameters as derived from the phase shifts of
\cite{Spi67,Boy72,Har72} in Fig.~\ref{fig:a7phase}. From the shown angular
distributions the parameter $\lambda^{\rm{even}}$ may vary between $1.40$ and
$1.45$; a careful inspection of the phase shift calculations in
Fig.~\ref{fig:a7phase} shows that the low-energy $s$-wave phase shifts of
\cite{Boy72} and the overall energy dependence are best described using
$\lambda^{\rm{even}} =1.45$; however, the data from \cite{Spi67} are better
desribed using $\lambda^{\rm{even}} =1.40$ (green dashed line in
Fig.~\ref{fig:a7phase}). Because of the better overall description of the
$s$-wave phase shifts and the excellent description of the $d$-wave phase
shifts I take $\lambda^{\rm{even}} =1.45$, $\lambda^{\rm{even}}_{LS} = 0$ as
the standard calculation in the following, keeping $\lambda^{\rm{odd}} =
1.830$ and $\lambda_{LS}^{\rm{odd}} = -0.173$ fixed for the odd partial
waves. An additional calculation with $\lambda^{\rm{even}} =1.45$,
$\lambda^{\rm{even}}_{LS} = -0.173$ (instead of $\lambda^{\rm{even}}_{LS} =
0$) is almost identical to the previous calculation in \cite{Mohr93} except
minor technical modifications (e.g.\ larger integration range and smaller step
size).

A minor disagreement between the calculated angular distribution and the
experimental data can be seen at $E = 2715$\,keV. This is not very surprising
because this energy is at the low-energy tail of the $7/2^-$ resonance. The
width of this resonance is slightly underestimated: the calculated width is
$\Gamma_{\rm{calc}} = 145$\,keV which has to be compared to the experimental
width of $\Gamma_{\rm{exp}} = 175 \pm 7$\,keV (see also Table
\ref{tab:level}). A much better agreement is obtained if the resonance
contribution is artificially enhanced in the calculation by an increase of the
phase shift of the $7/2^-$ partial wave by about $5^\circ$ from $10^\circ$ in
the standard calculation to $15^\circ$ (dotted black line in
Fig.~\ref{fig:scatt}). The deviation is clearly not related to the potential
strength of the even partial waves which will be most important for the
analysis of the \heag\ capture cross section in the next
Sec.~\ref{sec:rescapt}.

Summarizing the above, an excellent agreement between all experimental
scattering data (phase shifts and angular distributions) and the calculated
values is obtained using the above derived potential strength parameters
$\lambda^{\rm{even}} = 1.452$, $\lambda_{LS}^{\rm{even}} = 0$,
$\lambda^{\rm{odd}} = 1.830$ and $\lambda_{LS}^{\rm{odd}} = -0.173$ (standard
case). The results of all scattering experiments are consistent with
each other and can be well described using the present model.

The scattering wave functions can now be calculated without further adjustment
of parameters. Because of the dominating external capture (see
Fig.~\ref{fig:overlap}), good agreement between the predicted and experimental
capture cross sections has to be expected. This will be further analyzed in
the next Sec.~\ref{sec:rescapt}.

\subsection{The \heag\ capture cross section}
\label{sec:rescapt}
In addition to the scattering wave function $\chi(r)$, the determination of
the DC cross section in Eqs.~(\ref{eq:sig_th}) and (\ref{eq:DC}) requires the
calculation of the bound state wave function $u(r)$. Because of the
contributions from the nuclear exterior (see Fig.~\ref{fig:overlap}), it is
essential that the calculated bound state wave function has the correct
asymptotic behavior that is defined by the binding energy. This leads to a
slight readjustment of the potential strength parameter $\lambda$ for the
bound states compared to the above results for the scattering phase
shifts. The results are $\lambda(3/2^-) = 1.836$ for the ground state of $^7$Be
at $E = -1587$\,keV and $\lambda(1/2^-) = 1.799$ for the first excited state
of $^7$Be at $E = -1158$\,keV ($E^\ast = 429$\,keV). No spin-orbit potential
has been used here.

\begin{figure}[thbp] 
  \centering
  \includegraphics[bbllx=95,bblly=75,bburx=345,bbury=565,width=8.5cm]{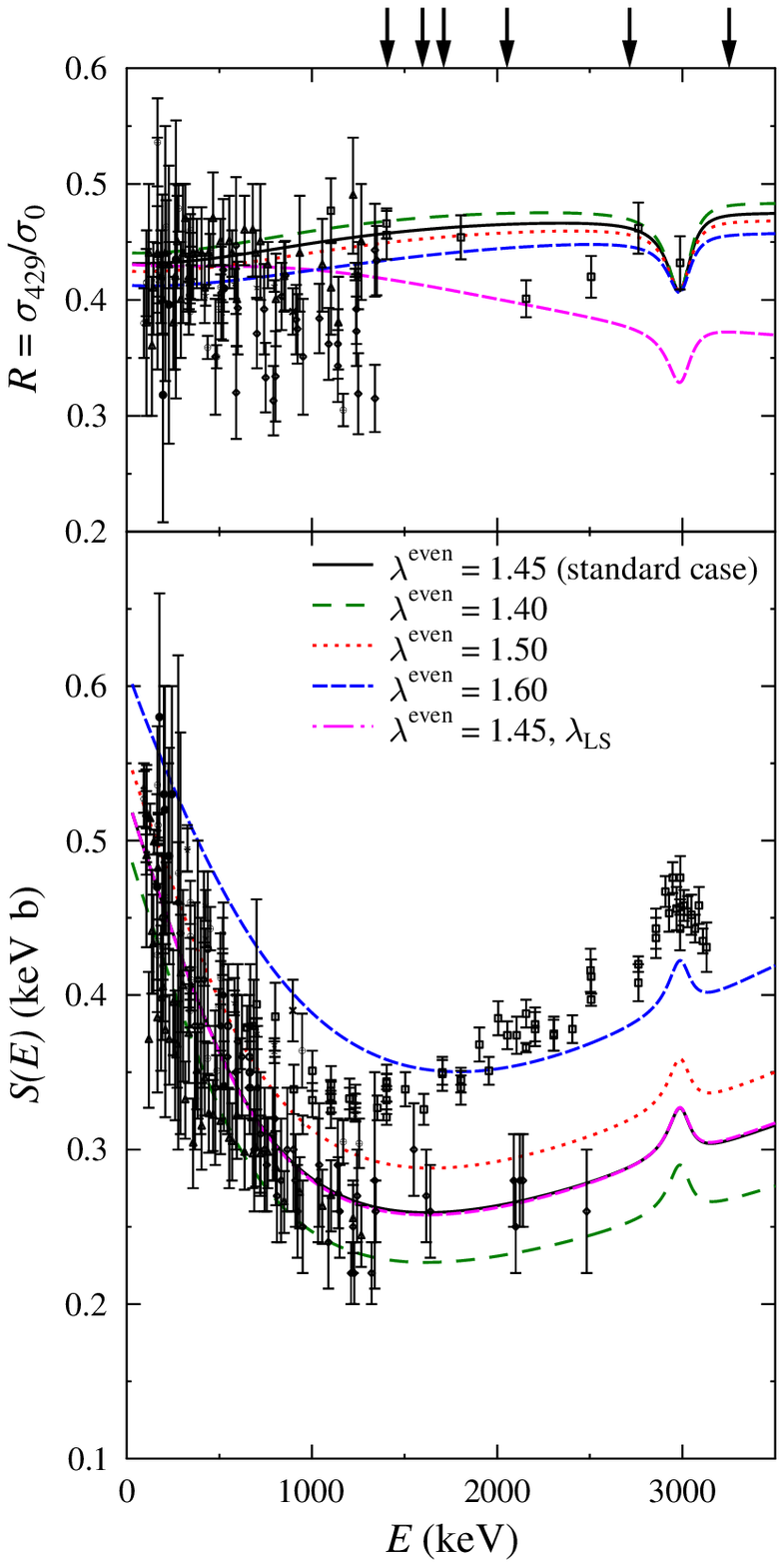}
  \caption{
    (Color online)
    Astrophysical \sfac\ and branching ratio $R = \sigma_{429}/\sigma_{0}$ for
    the \heag\ capture reaction. Experimental data are taken from
    \cite{Hol59,Par63,Nag69,Kra82,Rob83,Volk83,Alex84,Osb84,Hil88,Nar04,Bemm06,Gyu07,Bro07,Con07,Cos08,Lev09}. The
    linestyles and color codes are the same as in Figs.~\ref{fig:a7phase} and
    \ref{fig:scatt}. The vertical arrows on top indicate the energies of the
    angular distributions in the previous Fig.~\ref{fig:scatt} where the
    elastic scattering cross sections are well reproduced. These scattering
    data cover the energy range of the recent \heag\ capture data
    \cite{Lev09} (shown as open squares).
}
  \label{fig:sfact1}
\end{figure}

\begin{figure}[thbp] 
  \centering
  \includegraphics[bbllx=95,bblly=75,bburx=360,bbury=410,width=8.5cm]{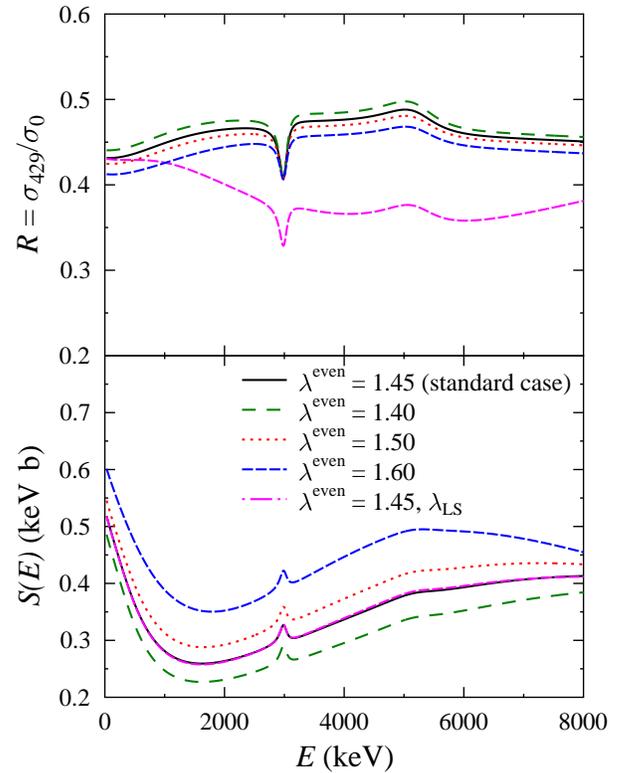}
  \caption{
   (Color online)
    Same as Fig.~\ref{fig:sfact1}, but with an extended energy range up to
    8\,MeV. For better visibility of the calculated curves, the experimental
    data have been omitted. The linestyles and color codes are the same as in
    the previous Figs.~\ref{fig:a7phase}, \ref{fig:scatt}, and
    \ref{fig:sfact1}. 
}
  \label{fig:sfact2}
\end{figure}

The remaining calculation of the DC cross section is straightforward and does
not require any adjustment to experimental capture data. The results are shown
in Fig.~\ref{fig:sfact1} with a comparison to the available experimental data
and in Fig.~\ref{fig:sfact2} in a broader energy range. The calculations take
into account all possible $E1$, $E2$, and $M1$ contributions from the incoming
scattering waves with $L = 0 - 3$. However, as has been shown in several
previous studies, the $E1$ contribution is dominating at all energies. Only in
the $7/2^-$ and $5/2^-$ resonances a noticeable $E2$ contribution is found,
and the $M1$ contribution remains negligible over the whole energy range. A
decomposition into the various contributions has already been shown in Fig.~5
of \cite{Tom63}, Fig.~8 of \cite{Kim81}, Fig.~8 of \cite{Liu81}, Fig.~2 of
\cite{Mer86}, Fig.~7 of \cite{Mohr93}, Fig.~2 of \cite{Dub95}, Fig.~8 of
\cite{Nol01}, and Fig.~5 of \cite{Mas09}, and is thus not repeated here.

The standard calculation (using the potential strength parameters from the
adjustment to the experimental phase shifts) shows an excellent agreement with
the total \sfac\ and the branching ratio $R = \sigma_{429}/\sigma_0$ in the
low-energy region below 1\,MeV. At higher energies the calculation agrees with
the old data of Parker and Kavanagh \cite{Par63}, but is significantly lower
than the recent experimental data of \cite{Lev09}.

The sensitivity of the \heag\ capture cross section on the underlying
potential has been studied in the same way as in the previous
Sec.~\ref{sec:resscat} by a variation of the $\lambda$ and $\lambda_{LS}$
potential strength parameters. The results are also shown in
Figs.~\ref{fig:sfact1} and \ref{fig:sfact2}. In general, an increasing
potential strength leads to an increasing \sfac . A detailed discussion will
be given in Sec.~\ref{sec:conc}.

Before entering the detailed discussion of the shown \heag\ capture cross
sections, the presented results are completed by an analysis of the two $L=3$
resonances and a study of the low-energy behavior of the astrophysical \sfac
. For both resonances it is obvious that the total width $\Gamma$ is
practically identical to the alpha particle width $\Gamma_\alpha$. As already
shown in Table \ref{tab:level}, the calculated widths $\Gamma_{\rm{calc}}$
agree reasonably well with the experimental results $\Gamma_{\rm{exp}}$. This
is also obvious from correct description of the resonant behavior of the $L =
3$ elastic phase shifts in Fig.~\ref{fig:a7phase}.

The $7/2^-$ resonance at 2983\,keV decays by an $E2$ transition to the $3/2^-$
ground state of $^7$Be whereas no $E2$ transition to the first ecited $1/2^-$
state is possible. A Breit-Wigner fit to the calculated capture cross section
in this resonance leads to a radiation width $\Gamma_{\gamma} =
\Gamma_{\gamma,0} = 37$\,meV and a resonance strength $\omega \gamma =
150$\,meV. This prediction is in reasonable agreement with the experimental
result of $\omega \gamma = 330 \pm 210$\,meV \cite{Lev09}.

The $5/2^-$ resonance at 5143\,keV was not measured yet in the
\heag\ reaction. The DC model predicts $\Gamma_{\gamma,0} \approx 50$\,meV for
the $E2$ transition to the $3/2^-$ ground state and $\Gamma_{\gamma,429}
\approx 185$\,meV for the $E2$ transition to the first excited $1/2^-$
state. This leads to $\Gamma_\gamma = \Gamma_{\gamma,0} + \Gamma_{\gamma,429}
\approx 235$\,meV or a total resonance strength $\omega \gamma \approx
0.7$\,eV. Because of the huge width of this resonance, any determination of
resonance parameters has relatively large uncertainties. An additional
theoretical uncertainty is related to the opening of the $^6$Li-$p$ channel
slightly above 4\,MeV.

It is interesting to note that the different decay patterns of the $7/2^-$ and
$5/2^-$ resonances are clearly visible in the branching ratio $R$ although the
$E2$ contribution to the total cross section is small compared to the
dominating $E1$ transitions. Fig.~\ref{fig:sfact2} clearly shows a smaller
branching ratio in the $7/2^-$ resonance around 3\,MeV which can only decay to
the $3/2^-$ ground state, and a larger branching ratio in the $5/2^-$
resonance around 5\,MeV which predominantly decays to the $1/2^-$ first
excited state.

The \sfac\ at very low energies is extrapolated down to $S(0)$ using the
following procedure. The calculated cross sections are fitted using a
second-order polynomial up to 500\,keV. The results for $S(0)$ and
$S'(0)/S(0)$ are listed in Table \ref{tab:S0} for the different potentials
shown in Figs.~\ref{fig:a7phase}, \ref{fig:scatt}, \ref{fig:sfact1}, and
\ref{fig:sfact2}. The standard case leads to $S(0) = 0.53$\,keV\,b and
$S'(0)/S(0) = -0.73$\,MeV$^{-1}$, in agreement with the compilation of solar
fusion cross sections by Adelberger {\it et al.}\ \cite{SolFus} who recommend
$S(0) = 0.53 \pm 0.05$\,keV\,b and $S'(0)/S(0) = -0.57$\,MeV$^{-1}$, and the
NACRE compilation \cite{NACRE} where $S(0) = 0.54 \pm 0.09$\,keV\,b and
$S'(0)/S(0) = -0.96$\,MeV$^{-1}$ are given. The recent summary of Cyburt and
Davids \cite{Cyb08} recommends a slightly higher value of $S(0) = 0.580 \pm
0.043$\,keV\,b and $S'(0)/S(0) = -0.92 \pm 0.18$\,MeV$^{-1}$. Variations of
the order of about 10\,\% for the slope of the \sfac\ at zero energy are found
using the different potentials. Such variations are also typical for the
different theoretical calculations in
\cite{Chr61,Tom61,Tom63,Kim81,Liu81,Wal83,Kaj84,Wal84,Mer86,Kaj87,Buck88,Mohr93,Dub95,Bay00,Cso00,Nol01,Des04,Can08,Mas09,Nav09}. The
adopted $S'(0)/S(0)$ \cite{SolFus,NACRE,Cyb08} show somewhat larger
variations. The average of the three compilations of $S'(0)/S(0) =
-0.82$\,MeV$^{-1}$ is in reasonable agreement with the calculated $S'(0)/S(0)
= -0.73$\,MeV$^{-1}$ in the standard case and in perfect agreement with the
calculation using $\lambda^{\rm{even}} = 1.40$ where $S'(0)/S(0) =
-0.79$\,MeV$^{-1}$ is found. However, larger values $\lambda^{\rm{even}} >
1.50$ correspond to lower $S'(0)/S(0)$.
\begin{table}[bthp]
\caption{
Extrapolated astrophysical \sfac\ $S(0)$ at zero energy and its normalized
derivative $S'(0)/S(0)$ for different potentials. The strength
parameters $\lambda^{\rm{odd}} = 1.830$ and $\lambda_{LS}^{\rm{odd}} =
-0.173$ have not been varied. For comparison, the results of recent
compilations \cite{SolFus,NACRE,Cyb08} are also listed.
\label{tab:S0}
}
\begin{center}
\begin{tabular}{ccccl}
\hline
  $\lambda^{\rm{even}}$ & $\lambda_{LS}^{\rm{even}}$ &
  $S(0)$ & $S'(0)/S(0)$ & remarks\\
 & & (keV\,b) & (MeV$^{-1}$) \\
\hline
1.45 & 0.0 & 0.530 & $-0.731$ & standard case\\
1.40 & 0.0 & 0.497 & $-0.791$ & \\
1.50 & 0.0 & 0.557 & $-0.666$ & \\
1.60 & 0.0 & 0.611 & $-0.536$ & \\
1.45 & -0.173 & 0.530 & $-0.733$ & close to Ref.~\cite{Mohr93}\\
\hline
$-$  & $-$ & $0.53 \pm 0.05$ & $-0.57$ & Adelberger {\it et al.}\ \cite{SolFus} \\
$-$  & $-$ & $0.54 \pm 0.09$ & $-0.96$ & NACRE \cite{NACRE} \\
$-$  & $-$ & $0.580 \pm 0.043$ & $-0.92$ & Cyburt and Davids \cite{Cyb08} \\
\hline
\end{tabular}
\end{center}
\end{table}

Using the standard case potential, $S(0) = 0.53$\,keV\,b is predicted from the
present study. This prediction depends on the theoretical spectroscopic
factors $C^2 S$ of the bound state wave functions. From the excellent
agreement between the experimental and calculated capture cross sections and
branching ratios at low energies (see Fig.~\ref{fig:sfact1}) it is obvious that
the chosen theoretical spectroscopic factors \cite{Kur75} are correct within
about 10\,\%. Any minor readjustment of these spectroscopic factors will
affect the the calculated capture cross sections linearly at all energies, but
will not affect the calculated energy dependence of the \heag\ capture cross
section. Also the normalized slope $S'(0)/S(0)$ will not be affected by a
readjustment of the spectroscopic factors.

The chosen DC model does not require effective charges and uses $q = +2e$ for
the involved $^3$He and $^4$He nuclei. An additional introduction of effective
charges in the model may affect the absolute values of the DC cross
sections. Because of the dominating E1 contribution to the capture cross
section, the introduction of effective charges does not affect the calculated
energy dependence in a significant way. Similar to the statements on the
spectroscopic factor in the previous paragraph, effective charges do also not
affect the normalized slope $S'(0)/S(0)$ of the astrophysical \sfac . Although
effective charges cannot be determined from scattering phase shifts, it is
obvious from the excellent agreement between the experimental and calculated
cross sections at low energies that the effective charges cannot deviate
strongly from the used real charges.

\section{Discussion and conclusions}
\label{sec:conc}
It has been shown in the previous Sects.~\ref{sec:resscat} and
\ref{sec:rescapt} that the standard calculation with the parameters
$\lambda^{\rm{even}} = 1.45$, $\lambda^{\rm{even}}_{LS} = 0$,
$\lambda^{\rm{odd}} = 1.83$, and $\lambda^{\rm{odd}}_{LS} = -0.173$ is able to
describe elastic scattering phase shifts for \heaa\ reaction in the full
energy range up to about 8\,MeV and the
\heag\ capture cross sections to the $^7$Be ground state and first excited
state at low energies. At higher energies the calculation agrees with the old
data \cite{Par63} but disagrees with the new data \cite{Lev09}. In the
following discussion the theoretical uncertainties will be analyzed, and the
question will be answered whether the new data of \cite{Lev09} can be
described within the present model. Most of the considerations will also be
valid for other models because of the dominating external contributions to the
\heag\ capture cross section. The discussion will focus on the dominating $E1$
contributions from the incoming $s$-wave and $d$-wave.

The analysis of elastic scattering clearly shows that the $s$-wave and
$d$-wave phase shifts can only be reproduced using the potential strength of
the standard calculation (see Figs.~\ref{fig:a7phase} and
\ref{fig:scatt}). For the $s$-wave $\lambda^{\rm{even}} = 1.40 - 1.45$ is the
acceptable range. The $d$-wave phase shifts which remain close to zero up to
about 8\,MeV can only be described with $\lambda$ values very close to
$\lambda^{\rm{even}} = 1.45$. A value of $\lambda^{\rm{even}} > 1.50$ can
clearly be excluded. This is further strengthened by the analysis of the slope
$S'(0)/S(0)$ at low energies, see Table \ref{tab:S0}.

The integrand of the DC integral in Eq.~(\ref{eq:DC}), i.e.\ the overlap
between scattering wave function $\chi(r)$, electromagnetic operator
${\cal{O}}^{E1}$, and bound state wave function $u(r)$, is shown for the
standard calculation in Fig.~\ref{fig:overlap}. The upper part shows the $E1$
transition from the $s$-wave to the ground state and first excited state, and
the middle part shows the $E1$ transition from the $d$-wave to the ground
state in $^7$Be. As usual, the integrand is shifted towards the exterior for
lower energies. But it has to be pointed out that all curves are clearly
dominated by contributions from far outside the nucleus at both considered
energies $E = 100$\,keV and 3\,MeV; i.e., also at the high energy $E = 3$\,MeV
the external contribution dominates. Thus, the result of the DC calculation is
essentially defined by the behavior of the scattering wave function $\chi(r)$
and the bound state wave function $u(r)$ at radii between 5\,fm and 35\,fm
that is much larger than the size of $^7$Be. Any calculation with realistic
nuclear potentials should provide a more or less similar capture cross section
if the potential is able to reproduce the elastic phase shifts. This is
particularly true for the calculated energy dependence.

The absolute value of the bound state wave function $u(r)$ in the external
region depends indirectly on the chosen potential shape. The slope of the
external part of the bound state wave function $u(r)$ is well defined by the
binding energy, but the normalization of the bound state wave function $u(r)$
\begin{equation}
\int_0^\infty u^2(r) \, dr = 1
\label{eq:norm}
\end{equation}
has its main contribution from smaller radii $r < 5$\,fm, i.e.\ from the
nuclear interior and surface (see Fig.~\ref{fig:bound}). In line with the
considerations in \cite{Ili08}, the chosen spectroscopic factors $C^2 S
\approx 1.2$ for the ground state and first excited state in $^7$Be simply
mean here that the bound state wave functions $u(r)$ in the nuclear exterior
have to be scaled by a factor $\sqrt{C^2 S} \approx 1.1$ to bring the
calculated DC cross section into agreement with the experimental data at low
energies. The choice of a different potential leads to a different shape of
the wave function $u(r)$ and its asymptotic behavior that can be compensated
by a different spectroscopic factor. Finally, the theoretical cross section,
in particular its energy dependence, will remain almost unaffected -- provided
that the phase shifts are well reproduced.

Nevertheless, an attempt has been made to reproduce the new \heag\ capture
data \cite{Lev09} within the present model. In general, an increasing nuclear
potential strength decreases the Coulomb barrier and thus increases the
calculated capture cross section. A value of $\lambda^{\rm{even}} \approx 1.6$
is required to reproduce the larger cross section data of \cite{Lev09}. The
calculated capture cross section at low energies is increased by about 15\,\%
compared to the standard case which is slightly higher than the allowed range
of the precision experiments at low energies. This larger potential strength
also leads to a strong disagreement with the scattering data. In particular,
the elastic scattering angular distributions in Fig.~\ref{fig:scatt} cannot be
described with the increased value of $\lambda^{\rm{even}} = 1.6$. These
angular distributions have been measured in an energy range that covers the
energy range of the recent \heag\ capture data \cite{Lev09} (indicated by the
arrows on top of Fig.~\ref{fig:sfact1}). Taking into account the above
considerations on the radial behavior of the overlap integral in
Eq.~(\ref{eq:DC}), it can be excluded that the DC model is able to reproduce
simultaneously \heaa\ elastic scattering \cite{Bar64,Mohr93,Spi67,Boy72,Har72}
and \heag\ capture \cite{Lev09} at energies around $2 -3$\,MeV.

It is interesting to note that the branching ratio $R = \sigma_{429}/\sigma_0$
is practically not affected by the variation of the potential strength of the
$s$- and $d$-waves because both transitions are increased in the same way when
$\lambda^{\rm{even}}$ is increased from its standard value of 1.45 to about
1.6 to fit the higher capture data of \cite{Lev09}. However, the branching
ratio $R$ is significantly affected by the spin-orbit potential of the
$d$-wave. Using the spin-orbit potential strength from the odd partial waves
also for the even partial waves ($\lambda^{\rm{even}}_{LS} = -0.173$ instead
of the standard case $\lambda^{\rm{even}}_{LS} = 0$), the effective strength
of the potential for the $d_{5/2}$ ($d_{3/2}$) wave is increased
(decreased). As pointed out above, an increased potential strength leads to
increased capture cross sections. Because there is only an $E1$ transition
from the $d_{5/2}$ wave to the $3/2^-$ ground state, but no $E1$ transition to
the $1/2^-$ first excited state, the ground state contribution is enhanced,
and thus the branching ratio $R$ decreases at higher energies (see
Figs.~\ref{fig:sfact1} and \ref{fig:sfact2}). However, at low energies the
dominating contribution to the capture cross section comes from the $s$-wave
which is not affected by the spin-orbit potential, and the branching ratio $R$
at very low energies is practically identical to the standard case.

It is difficult to find an explanation for the discrepancy between the
calculated \heag\ capture cross sections and the recent experimental data
\cite{Lev09}. Experimental scattering data have been obtained in a series of
independent experiments \cite{Bar64,Spi67,Boy72,Har72,Iva68,Chu71,Mohr93} that
agree well with each other. The determination of the scattering wave function
in the DC integral in Eq.~(\ref{eq:DC}) is thus very well-defined. Although
the requested accuracy of a few per cent for astrophysical modelling has not
yet been achieved for the \heag\ capture reaction, all experimental capture
data at low energies agree within each other at a $10 - 15$\,\% level. The
spectroscopic factors -- taken as normalization factors in Eq.~(\ref{eq:DC})
-- are thus also well-defined at this level. The same level of accuracy should
then be expected for the energy range of the new \heag\ capture data
\cite{Lev09} up to about 3\,MeV. Instead, the standard calculation
underestimates the new capture data by about $30 - 40$\,\%. It has to be
pointed out that a wrong normalization of the new data \cite{Lev09} is also
very unlikely because the experiment has studied the \heag\ capture reaction
very carefully using three different independent techniques (recoil detection,
$\gamma$ detection, and activation) that agree very well with each other at
the level of a few per cent. Of course, an independent confirmation of the new
data in \cite{Lev09} is nevertheless highly desireable, in particular, because
of the discrepancy between the old data in \cite{Par63} and the new data in
\cite{Lev09}.

The only remaining explanation is that the DC model itself is not appropriate
for the description of the \heag\ capture reaction. However, even ``exotic''
theoretical solutions like other electromagnetic transitions (except $E1$,
$E2$, and $M1$) or further unobserved bound states in $^7$Be can be excluded
at the requested level of $30 - 40$\,\% from the agreement of the different
experimental techniques in \cite{Lev09}. And because of the dominating exteral
contributions to the DC cross section, any theoretical calculation that
reproduces the scattering phase shifts, should provide a similar theoretical
energy dependence and thus underestimate the new capture data for the
\heag\ reaction in \cite{Lev09} in a similar way.

\section{Summary}
\label{sec:summ}
The \heaa\ elastic scattering and \heag\ capture cross sections have been
studied within the DC model together with folding potentials. It is shown that
the elastic scattering phase shifts are well reproduced over a broad energy
range. Whereas the low-energy capture data are well described simultaneously
with the phase shifts, there is a significant underestimation of the recent
capture data \cite{Lev09}, but agreement with the old data of \cite{Par63}. As
the DC calculation is very well constrained by experimental scattering data
over the whole energy range studied in \cite{Lev09}, it is difficult to find
an explanation for the discrepancy between the calculated and the recently
measured capture cross section of the \heag\ reaction. Further eperimental and
theoretical effort is needed to resolve this surprising problem.

\begin{acknowledgments}
Encouraging discussions and all the support in reconstructing old experimental
data is gratefully acknowledged - many thanks to  A.\ Barnard, R.\ Cyburt,
B.\ Davids, P.\ Descouvemont, T.\ Hehl, and F.\ Strieder.
\end{acknowledgments}


\begin{thebibliography}{99}
%
\bibitem{Cyb04}
R.\ H.\ Cyburt, B.\ D.\ Fields, K.\ A.\ Olive,
\prd {\bf 69}, 123519 (2004).
%
\bibitem{Lev09}
A.\ Di Leva, L.\ Gialanella, R.\ Kunz, D.\ Rogalla, D.\ Sch\"urmann,
F.\ Strieder, M. De Cesare, N.\ De Cesare, A.\ D'Onofrio, Zs.\ F\"ul\"op,
Gy.\ Gy\"urky, G.\ Imbriani, G.\ Mangano, A.\ Ordine, V.\ Roca, C.\ Rolfs,
M.\ Romano, E.\ Somorjai, F.\ Terrasi,
\prl , submitted.
%
\bibitem{Hol59}
H.\ D.\ Holmgren and R.\ L.\ Johnston,
Phys.\ Rev.\ {\bf 113}, 1556 (1959).
%
\bibitem{Par63}
P.\ Parker and R.\ Kavanagh,
Phys.\ Rev.\ {\bf 131}, 2578 (1963).
%
\bibitem{Nag69} 
K.~Nagatani, M.~R.~Dwarakanath, D.~Ashery,
Nucl.\ Phys.\ {\bf A128}, 325 (1969).
%
\bibitem{Kra82} 
H.~Kr\"awinkel, H.~W.~Becker, L.~Buchmann, J.~G\"orres, K.~U.~Kettner,
W.~E.~Kieser, R.~Santo, P.~Schmalbrock, H.~P.~Trautvetter, A.~Vlieks,
C.~Rolfs, J.~W.~Hammer, R.~E.~Azuma, W.~S.~Rodney,
Z.\ Phys.\ A {\bf 304}, {307} (1982).
%
\bibitem{Rob83}
R.\ G.\ H.\ Robertson, P.\ Dyer, T.\ J.\ Bowles, R.\ E.\ Brown, N.\ Jarmie,
C.\ J.\ Maggiore, S.\ M.\ Austin,
\prc {\bf 27}, 11 (1983).
%
\bibitem{Volk83}
H.\ Volk, H.\ Kr\"awinkel, R.\ Santo, L. Wallek,
Z.\ Phys.\ A {\bf 310}, 91 (1983).
%
\bibitem{Alex84}
T.\ K.\ Alexander, G.\ C.\ Ball, W.\ N.\ Lennard, H.\ Geissel, H.-B.\ Mak,
Nucl.\ Phys.\ {\bf A427}, 526 (1984).
%
\bibitem{Osb84} 
J.~L.~Osborne, C.~A.~Barnes, R.~W.~Kavanagh, R.~M.~Kremer, G.~J.~Mathews,
J.~L.~Zyskind, P.~D.~Parker, A.~J.~Howard, 
\prl {\bf 48}, {1664} (1982); Nucl.\ Phys.\ {\bf A419}, {115} (1984).
%
\bibitem{Hil88} M.~Hilgemeier, H.~W.~Becker, C.~Rolfs, H.~P.~Trautvetter,
  J.~W.~Hammer, 
Z.\ Phys.\ A {\bf 329}, {243} (1988).
%
\bibitem{Nar04}
B.\ S.\ Nara Singh, M.\ Hass, Y.\ Nir-El, G.\ Haquin,
\prl {\bf 93}, 262503 (2004).
%
\bibitem{Bemm06}
D.\ Bemmerer, F.\ Confortola, H.\ Costantini, A.\ Formicola, Gy.\ Gy\"urky,
R.\ Bonetti, C.\ Broggini, P.\ Corvisiero, Zs.\ Elekes, Zs.\ F\"ul\"op,
G.\ Gervino, A.\ Guglielmetti, C.\ Gustavino, G.\ Imbriani, M.\ Junker,
M.\ Laubenstein, A.\ Lemut, B.\ Limata, V.\ Lozza, M.\ Marta, R.\ Menegazzo,
P.\ Prati, V.\ Roca, C.\ Rolfs, C.\ Rossi Alvarez, E.\ Somorjai,
O.\ Straniero, F.\ Strieder, F.\ Terrasi, H.\ P.\ Trautvetter, 
\prl {\bf 97}, 122502 (2006).
%
\bibitem{Gyu07}
Gy.\ Gy\"urky, F.\ Confortola, H.\ Costantini, A.\ Formicola, D.\ Bemmerer,
R.\ Bonetti, C.\ Broggini, P.\ Corvisiero, Zs. Elekes, Zs.\ F\"ul\"op,
G.\ Gervino, A.\ Guglielmetti, C.\ Gustavino, G.\ Imbriani, M.\ Junker,
M.\ Laubenstein, A.\ Lemut, B.\ Limata, V.\ Lozza, M.\ Marta, R.\ Menegazzo,
P.\ Prati, V.\ Roca, C.\ Rolfs, C.\ Rossi Alvarez, E.\ Somorjai,
O.\ Straniero, F.\ Strieder, F.\ Terrasi, H.\ P.\ Trautvetter,
\prc {\bf 75}, 035805 (2007).
%
\bibitem{Con07}
F.\ Confortola, D.\ Bemmerer, H.\ Costantini, A.\ Formicola, Gy.\ Gy\"urky,
P.\ Bezzon, R.\ Bonetti, C.\ Broggini, P.\ Corvisiero, Zs.\ Elekes,
Zs.\ F\"ul\"op, G.\ Gervino, A.\ Guglielmetti, C.\ Gustavino, G.\ Imbriani,
M.\ Junker, M.\ Laubenstein, A.\ Lemut, B.\ Limata, V.\ Lozza, M.\ Marta,
R.\ Menegazzo, P.\ Prati, V.\ Roca, C.\ Rolfs, C.\ Rossi Alvarez,
E.\ Somorjai, O.\ Straniero, F.\ Strieder, F.\ Terrasi, H.\ P.\ Trautvetter,
\prc {\bf 75}, 065803 (2007).
%
\bibitem{Bro07}
T.\ A.\ D.\ Brown, C.\ Bordeanu, K.\ A.\ Snover, D.\ W.\ Storm, D.\ Melconian,
A.\ L.\ Sallaska, S.\ K.\ L.\ Sjue, S.\ Triambak,
\prc {\bf 76}, 055801 (2007).
%
\bibitem{Cos08}
H.\ Costantini, D.\ Bemmerer, F.\ Confortola, A.\ Formicola, Gy.\ Gy\"urky,
P.\ Bezzon, R.\ Bonetti, C.\ Broggini, P.\ Corvisiero, Zs.\ Elekes,
Zs.\ F\"ul\"op, G.\ Gervino, A.\ Guglielmetti, C.\ Gustavino, G.\ Imbriani,
M.\ Junker, M.\ Laubenstein, A.\ Lemut, B.\ Limata, V.\ Lozza, M.\ Marta,
R.\ Menegazzo, P.\ Prati, V.\ Roca, C.\ Rolfs, C.\ Rossi Alvarez,
E.\ Somorjai, O.\ Straniero, F.\ Strieder, F.\ Terrasi, H.\ P.\ Trautvetter,
Nucl.\ Phys.\ {\bf A814}, 144 (2008).
%
\bibitem{SolFus}
E.\ G.\ Adelberger, S.\ M.\ Austin, J.\ N.\ Bahcall, A.\ B.\ Balantekin,
G.\ Bogaert, L.\ S.\ Brown, L.\ Buchmann, F.\ E.\ Cecil, A.\ E.\ Champagne,
L.\ de Braeckeleer, C.\ A.\ Duba, S.\ R.\ Elliott, S.\ J.\ Freedman, M.\ Gai,
G.\ Goldring, C.\ R.\ Gould, A.\ Gruzinov, W.\ C.\ Haxton, K.\ M.\ Heeger,
E.\ Henley, C.\ W.\ Johnson, M.\ Kamionkowski, R.\ W.\ Kavanagh,
S.\ E.\ Koonin, K.\ Kubodera, K.\ Langanke, T.\ Motobayashi,
V.\ Pandharipande, P.\ Parker, R.\ G.\ H.\ Robertson, C.\ Rolfs,
R.\ F.\ Sawyer, N.\ Shaviv, T.\ D.\ Shoppa, K.\ A.\ Snover, E.\ Swanson,
R.\ E.\ Tribble, S.\ Turck-Chieze, J.\ F.\ Wilkerson,
Rev.\ Mod.\ Phys.\ {\bf 70}, 1265 (1998).
%
\bibitem{NACRE}
C.\ Angulo, M.\ Arnould, M.\ Rayet, P.\ Descouvemont, D.\ Baye,
C.\ Leclercq-Willain, A.\ Coc, S.\ Barhoumi, P.\ Aguer, C.\ Rolfs, R.\ Kunz,
J.\ W.\ Hammer, A.\ Mayer, T.\ Paradellis, S.\ Kossionides, C.\ Chronidou,
K.\ Spyrou, S.\ Degl'Innocenti, G.\ Fiorentini, B.\ Ricci, S.\ Zavatarelli,
C.\ Providencia, H.\ Wolters, J.\ Soares, C.\ Grama, J.\ Rahighi, A.\ Shotter,
M.\ Lamehi Rachti,
Nucl.\ Phys.\ {\bf A656}, 3 (1999).
%
\bibitem{Chr61}
R.\ F.\ Christy and I.\ Duck,
Nucl.\ Phys.\ {\bf 24}, 89 (1961).
%
\bibitem{Tom61}
T.\ A.\ Tombrello and G.\ C.\ Phillips,
Phys.\ Rev.\ {\bf 122}, 224 (1961).
%
\bibitem{Tom63}
T.\ A.\ Tombrello and P.\ D.\ Parker,
Phys.\ Rev.\ {\bf 131}, 2582 (1963).
%
\bibitem{Kim81}
B.\ T.\ Kim, T.\ Izumoto, K.\ Nagatani,
\prc {\bf 23}, 33 (1981).
%
\bibitem{Liu81}
Q.\ K.\ K.\ Liu, H.\ Kanada, Y.\ C.\ Tang,
\prc {\bf 23}, 645 (1981).
%
\bibitem{Wal83}
H.\ Walliser, Q.\ K.\ K.\ Liu, H.\ Kanada, Y.\ C.\ Tang,
\prc {\bf 28}, 57 (1983).
%
\bibitem{Mer86}
T.\ Mertelmeier and H.\ M.\ Hofmann,
Nucl.\ Phys.\ {\bf A459}, 387 (1986).
%
\bibitem{Kaj84}
T.\ Kajino and A.\ Arima,
\prl {\bf 52}, 739 (1984).
%
\bibitem{Wal84}
H.\ Walliser, H.\ Kanada, Y.\ C.\ Tang,
\prl {\bf 53}, 399 (1984).

\bibitem{Kaj87}
T.\ Kajino, H.\ Toki, S.\ M.\ Austin,
\apj {\bf 319}, 531 (1987).
%
\bibitem{Buck88}
B.\ Buck and A.\ C.\ Merchant,
J.\ Phys.\ G {\bf 14}, L211 (1988).
%
\bibitem{Mohr93}
  P.~Mohr, H.~Abele, R.~Zwiebel, G.~Staudt, H.~Krauss, H.~Oberhummer,
  A.~Denker, J.~W.~Hammer, G.~Wolf,
  \prc {\bf 48} {1420} (1993).
%
\bibitem{Dub95}
S.\ B.\ Dubovichenko and A.\ V.\ Dzhazairov-Kakhramanov,
Phys.\ Atom.\ Nucl.\ {\bf 58}, 579 (1995); Yad.\ Fiz.\ {\bf 58}, 635 (1995);
{\it{nucl-th/9802080}}.
%
\bibitem{Bay00}
D.\ Baye and E.\ Brainis,
\prc {\bf 61}, 025801 (2000).
%
\bibitem{Cso00}
A.\ Cs\'ot\'o and K.\ Langanke,
Few Body Systems {\bf 29}, 121 (2000).
%
\bibitem{Nol01}
K.\ Nollett,
\prc {\bf 63}, 054002 (2001).
%
\bibitem{Des04}
P.\ Descouvemont, A.\ Adahchour, C.\ Angulo, A.\ Coc, E.\ Vangioni-Flam,
At.\ Data Nucl.\ Data Tables {\bf 88}, 203 (2004).
%
\bibitem{Can08}
L.\ Canton and L.\ G.\ Levchuk,
Nucl.\ Phys.\ {\bf A808}, 192 (2008).
%
\bibitem{Mas09}
A.\ Mason, R.\ Chatterjee, L.\ Fortunato, A.\ Vitturi,
Europ.\ Phys.\ J.\ A {\bf 39}, 107 (2009).
%
\bibitem{Nav09}
P.\ Navratil, S.\ Quaglioni, I.\ Stetcu, B.\ R.\ Barrett,
J.\ Phys.\ G {\bf 36}, 083101 (2009).
%
\bibitem{Cyb08}
R.\ H.\ Cyburt and B.\ Davids,
\prc {\bf 78}, 064614 (2008).
%
\bibitem{Spi67}
R.\ Spiger and T.\ A.\ Tombrello,
Phys.\ Rev.\ {\bf 163}, 964 (1967).
%
\bibitem{Boy72}
W.\ R.\ Boykin, S.\ D.\ Baker, D.\ M.\ Hardy,
Nucl.\ Phys.\ {\bf A195}, 241 (1972).
%
\bibitem{Har72}
D.\ M.\ Hardy, R.\ J.\ Spiger, S.\ D.\ Baker, Y.\ S.\ Chen, T.\ A\ Tombrello,
Nucl.\ Phys.\ {\bf 195}, 250 (1972).
%
\bibitem{Bar64}
A.\ C.\ L.\ Barnard, C.\ M.\ Jones, G.\ C.\ Phillips,
Nucl.\ Phys.\ {\bf 50}, 629 (1964).
%
\bibitem{Iva68}
M.\ Ivanovich, P.\ G.\ Young, G.\ G.\ Ohlsen,
Nucl.\ Phys.\ {\bf A110}, 441 (1968).
%
\bibitem{Chu71}
L.\ S.\ Chuang,
Nucl.\ Phys.\ {\bf A174}, 399 (1971).
%
\bibitem{Til02}
D.\ R.\ Tilley, C.\ M.\ Cheves, J.\ L.\ Godwin, G.\ M.\ Hale, H.\ M.\ Hofmann,
J.\ H.\ Kelley, C.\ G.\ Sheu, H.\ R.\ Weller,
Nucl.\ Phys.\ {\bf A708}, 3 (2002).
%
\bibitem{Kim87}
K.\ H.\ Kim, M.\ H.\ Park, B.\ T.\ Kim,
\prc {\bf 35}, 363 (1987).
%
\bibitem{Kur75}
D.\ Kurath and D.\ J.\ Millener,
Nucl.\ Phys.\ {\bf A238}, 269 (1975).
%
\bibitem{Ili08}
C.\ Iliadis, C.\ Angulo, P.\ Descouvemont, M.\ Lugaro, P.\ Mohr,
\prc {\bf 77}, 045802 (2008).
%
\bibitem{Muk08}
A.\ M.\ Mukhamedzhanov, F.\ M.\ Nunes, P.\ Mohr,
\prc {\bf 77}, 051601(R) (2008).
%
\bibitem {Sat79} 
  G.\ R.\ Satchler and W.\ G.\ Love,
  Phys.\ Rep.\ {\bf 55}, 183 (1979).
%
\bibitem{Kob84} 
  A.\ M.\ Kobos, B.\ A.\ Brown, R.\ Lindsay, and 
  R.\ Satchler,
  Nucl.\ Phys.\ {\bf A425}, 205 (1984).
%
\bibitem{Vri87}
H.\ de Vries, C.\ W.\ de Jager, and C.\ de Vries,
At.\ Data Nucl.\ Data Tables {\bf 36}, 495 (1987).
%
\bibitem{Wal85}
H.\ Walliser and T.\ Fliessbach,
\prc {\bf 31}, 2242 (1985).
%
\bibitem{Buck85}
B.\ Buck, R.\ A.\ Baldock, J. Alberto-Rubio,
J.\ Phys.\ G {\bf 11}, L11 (1985).
%
\bibitem{Kel87}
J.\ H.\ Kelley, D.\ R.\ Tilley, H.\ R.\ Weller, H.\ H.\ Hasan,
Nucl.\ Phys.\ {\bf A474}, 1 (1987).
%
\bibitem{Mohr94}  
P.~Mohr, V.~K\"olle, S.~Wilmes, U.~Atzrott, G.~Staudt, J.~W.~Hammer,
H.~Krauss, H.~Oberhummer,
\prc {\bf 50}, {1543} {(1994)}.
%
\end{thebibliography}
\end{document}